\documentclass{PoS}

\usepackage{amsmath,amssymb,mathtools}
\usepackage{enumerate}
\usepackage{booktabs}
\usepackage{multirow}
\usepackage{graphicx}
\usepackage{xspace}
\usepackage{subcaption}

\DeclareRobustCommand{\rT}{\mathrm{T}}

\DeclareRobustCommand{\pt}{\ensuremath{p_\rT}\xspace}
\DeclareRobustCommand{\ptj}[1]{\ensuremath{p_{\rT,#1}}\xspace}

\DeclareRobustCommand{\htp}{\ensuremath{\hat{H}_\rT}\xspace} 
\DeclareRobustCommand{\nnlojet}{NNLO\scalebox{0.8}{JET}\xspace}

\title{Jet cross sections at the LHC with \nnlojet}

\ShortTitle{Jet cross sections at the LHC with \nnlojet}

\author{James Currie, Nigel Glover \\
Institute for Particle Physics Phenomenology, Department of Physics, University of Durham, Durham, DH1 3LE, UK \\
E-mail: \email{james.currie@durham.ac.uk}, \email{e.w.n.glover@durham.ac.uk}}

\author{\speaker{Aude Gehrmann-De Ridder} \\
Institute for Theoretical Physics, ETH, CH-8093 Z\"urich, Switzerland \\
Physik-Institut, Universit\"at Z\"urich, Winterthurerstrasse 190, CH-8057 Z\"urich, Switzerland\\
E-mail: \email{gehra@phys.ethz.ch}}

\author{Thomas Gehrmann \\
Physik-Institut, Universit\"at Z\"urich, Winterthurerstrasse 190, CH-8057 Z\"urich, Switzerland\\
E-mail: \email{thomas.gehrmann@uzh.ch}}

\author{Alexander Huss\\
Theoretical Physics Department, CH-1211 Geneve 23, Switzerland\\
E-mail: \email{alexander.huss@cern.ch}}

\author{Jo\~{a}o Pires\\
Centro de Fisica Teorica de Particulas - CFTP, 
Instituto Superior Tecnico IST,
Universidade de Lisboa, Av. Rovisco Pais,
P-1049-001 Lisboa, Portugal\\
E-mail:\email{joao.ramalho.pires@tecnico.ulisboa.pt}}
  
\abstract{We review the status of NNLO calculations for jet cross sections at the LHC. 
In particular, we describe how perturbative stability and convergence can 
be used as criteria to select the most appropriate scales in the theoretical description 
of di-jet and single jet inclusive production.}

\FullConference{Loops and Legs in Quantum Field Theory (LL2018)\\
		29 April 2018 - 04 May 2018\\
		St. Goar, Germany}

\begin{document}

\section{Introduction}
Due to their large production rate, low-multiplicity jet cross sections are among the best-measured 
collider observables.
At the LHC, they are measured differentially in the jet kinematics with an accuracy often reaching the percent level 
~\cite{ATLAS7TEV, CMS7TEV, ATLAS8TEV, CMS8TEV, ATLAS13TEV, CMS13TEV}.
Combined with perturbative QCD predictions of comparable accuracy,  these precision 
data have the potential to determine 
fundamental parameters of the Standard Model with percent level errors 
and to constrain beyond Standard Model physics searches.
Those accurate measurements also provide an ideal ground for probing the perturbative behaviour of QCD predictions 
computed at a given order in the strong coupling $\alpha_s$. 

Theoretical predictions for hadron collider observables are built from two main constituents: the parton-level cross sections and the parton distribution functions. 
These predictions have two types of uncertainties: parametric and perturbative.
The parametric uncertainties arise from the ingredients to the predictions that cannot be computed from 
first principles, but are extracted from data: $\alpha_s$ and the parton distributions. 
They both depend on auxiliary scales: the renormalization scale ($\mu_R$) and the mass factorization scale ($\mu_F$). 

The perturbative uncertainty, which is our main concern here, arises from the truncation of the perturbative series and 
is most often quantified by varying the renormalisation and factorisation scales around some predefined common central value, 
referred to as central reference scale. 
This commonly used procedure provides an estimation of the theoretical uncertainty related to the unknown missing higher orders 
terms in the theoretical predictions. The choice of this central scale, although usually physically motivated, is arbitrary:
any suitable choice is a priori equally valid. 
 
Up to now, jet observables have been compared to data at NLO level only \cite{ellis,giele,nagy}. 
Most recently, di-jet \cite{dijets} and single jet inclusive \cite{singlejetPRL} production cross sections 
and related distributions have been computed to NNLO in QCD, including all 
partonic channels, but restricting the NNLO corrections to the numerically dominant leading colour and leading $N_F$ terms. 

These computations were performed using the \nnlojet framework, which is a parton-level event generator including all 
partonic channels relevant at a given order, and which provides the full kinematical information on all final state particles. 
The \nnlojet framework is a common infrastructure used to compute NNLO corrections for jet production processes, 
employing the antenna subtraction method \cite{ourant, currie} to capture all infrared divergencies from the corresponding matrix elements.  
A detailed description of the \nnlojet framework including the specific processes 
which have been implemented in this infrastructure up to now, can be found in \cite{Radcor}.
 
These NNLO calculations, as they include the knowledge of three orders in the perturbative expansion in $\alpha_s$ 
provide a unique opportunity to test expectations regarding the perturbative convergence and stability of theoretical predictions when higher order corrections are included in the computation of these observables.   
 
It was furthermore observed that at NNLO, 
different, but equally motivated choices of the central scale value resulted in substantially different predictions 
for jet observables yielding a significant perturbative uncertainty and preventing so far the use of jet data in PDF fits, see  \cite{Harland-Lang:2017ytb}. 

It is the purpose of this talk to present detailed studies on the impact 
of choosing a particular functional form for the central scales in di-jet and single jet inclusive production.

 \section{The di-jet production cross section}
\label{sec:dijets}

Di-jet observables are defined through the two jets with the largest transverse momentum in the event. 
In \cite{dijets} we have presented the NNLO calculation of di-jet production 
doubly differential in the invariant mass $m_{jj}$ and half of the absolute rapidity difference of the two leading jets, ($|y^*|=|y_{j1}-y_{j2}|/2$ ) and compared it to the available ATLAS data \cite{ATLAS7TEV} at 7 TeV. This comparison was performed using 
the following central scale choices: 
the invariant mass $m_{jj}$ and the average transverse momentum of the two leading jets, $\langle p_T \rangle$.

\begin{figure*}[t]
  \centering
    \includegraphics[width=0.45\textwidth]{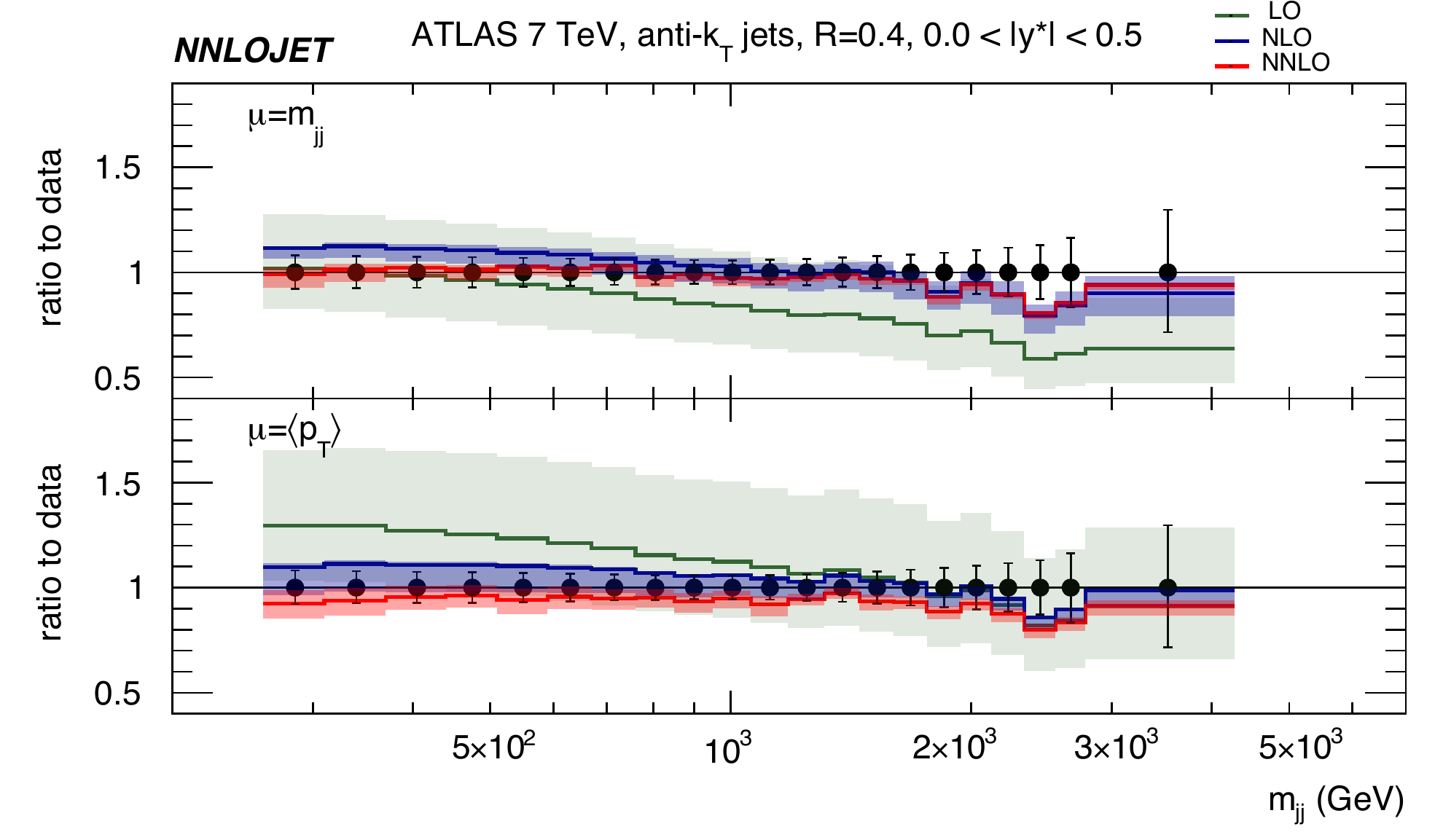}\hspace{1cm}
    \includegraphics[width=0.45\textwidth]{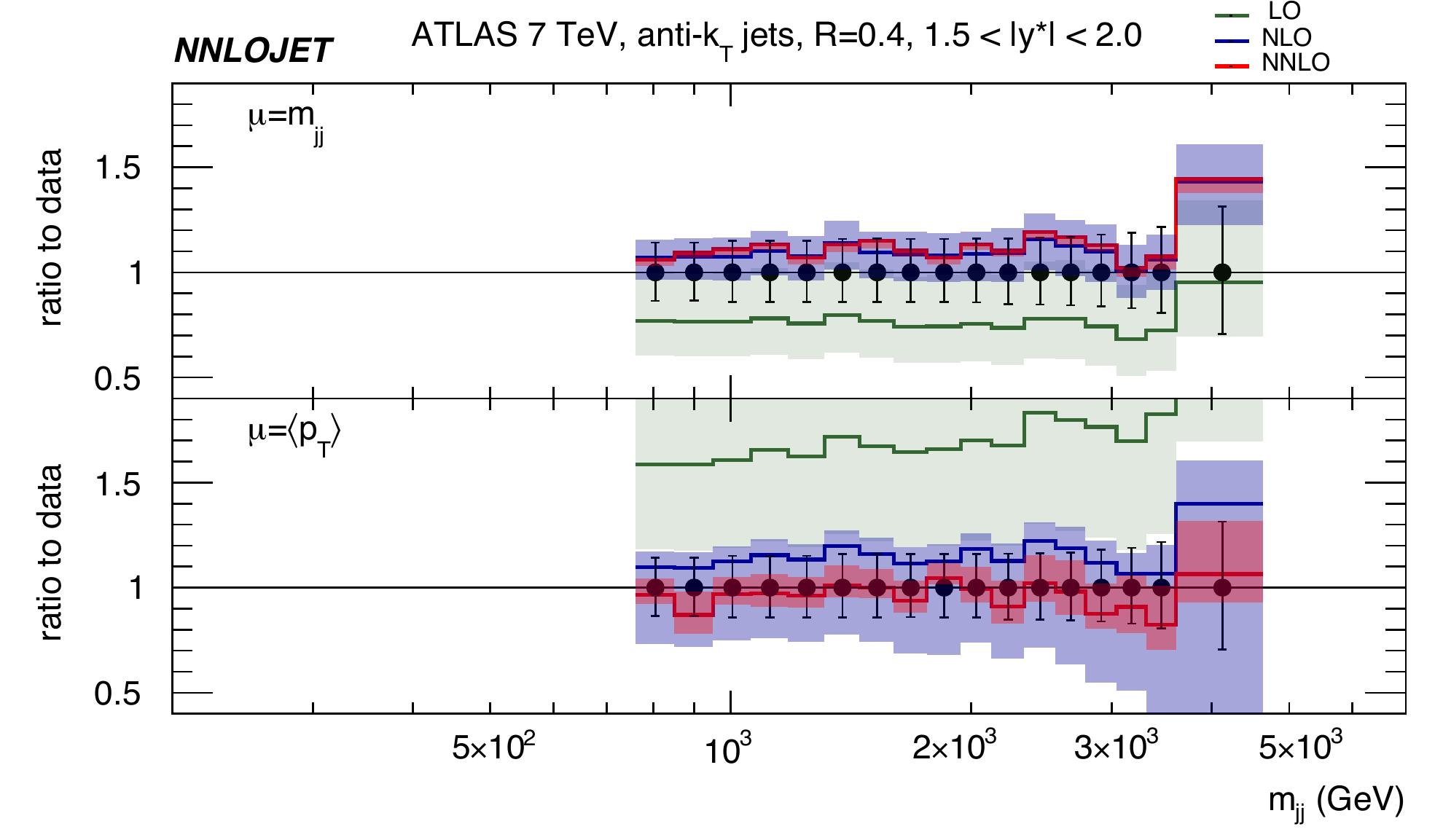}
  \caption{Ratio of theory predictions to data for $0.0<|y^*|<0.5$ (left) and $1.5<|y^*|<2.0$ (right) for the scale choices $\mu=m_{jj}$ (top) and $\mu=\langle p_{T}\rangle$ (bottom) at LO (green), NLO (blue) and NNLO (red). Scale bands represent variation of the cross section by varying the scales independently by factors of 2 and 0.5. \label{fig:dijets}}
 \end{figure*}

At NLO, it was previously noted that the predictions obtained with these two central scale choices are 
substantially different. This observation raised doubts on the reliability of the perturbative description of di-jet production, 
and the corresponding data were often not included in global determinations of parton distributions. 
Including the NNLO corrections, this spread in the predictions is reduced substantially, as can be seen in 
 Fig.\ref{fig:dijets}.  The spread is rapidity dependent, and 
 some impact of the central scale choice is still observed in the rapidity region $1.5<|y^*|<2.0$, while being largely 
 absent at low rapidity separation. 
While the spread is reduced considerably at NNLO, yielding mutually compatible predictions, 
the two central scale choices display a considerably different perturbative convergence and residual scale uncertainty. 
This leads us to conclude that the scale choice $\mu=m_{jj}$ appears to be 
 the most appropriate in the theoretical description of the di-jet cross section. 

With this choice as reference central scale, when comparing  
our NNLO predictions for the double differential cross section with the ATLAS data at 7 TeV \cite{ATLAS7TEV} ,
we found that those predictions yield a good agreement in shape and normalization with the data for all the 
kinematical range in invariant mass and rapidity. The inclusion of NNLO corrections  leads to
a significant improvement in the description of the data even for the low invariant mass and 
rapidity range where the NLO prediction deviated from it.  
It is also found that the residual scale uncertainty at NNLO is smaller than 
the experimental uncertainty for this observable.

These findings regarding the scale sensitivity in the di-jet observables cannot be transferred straightforwardly to the single jet inclusive production cross section, as we will see below.

\section{The single jet inclusive production cross section} 
The single jet inclusive production cross section is obtained by summing over all reconstructed jets in an event. By ordering 
the jets in transverse momentum, this cross section can be expressed as the sum of the 
jet cross sections for the first (leading), second, third (an so on) jets in the event. It 
has been studied extensively by the ATLAS and CMS collaborations 
~\cite{ATLAS7TEV, CMS7TEV, ATLAS8TEV, CMS8TEV, ATLAS13TEV, CMS13TEV}
as a function of the 
transverse momentum $p_{T}$ and rapidity $y$ 
at various center-of-mass energies ranging from $\sqrt{s}=7$~TeV  to $\sqrt{s}=13$~TeV. 

Beyond the requirement of observing at least one jet, no further constraints on the final state particles are imposed on this observable. 
As a consequence, an event can contain multiple jets and all jets that pass the jet fiducial cuts contribute individually to the cross section. 
Unlike in di-jet production where in a given distribution each event is counted once in a given kinematical bin, 
for the single jet inclusive production cross section, a single event can have multiple entries in the binned histogram. 

The phenomenological analysis of single jet inclusive processes 
thus turns out to be much more involved than in the di-jet production case. 
The fact that the contributions to inclusive distributions come from individual jets rather than events introduces 
more possibilities for  the
 choice of the central scale  in the theoretical 
predictions. We investigated this in detail in \cite{jetscales}, considering 
the following options (and multiples thereof):
\begin{itemize}
  \item  the individual jet transverse momentum \pt
  \item the leading-jet transverse momentum \ptj1 
   \item the scalar sum of the transverse momenta of all partons \htp
\end{itemize}
We can distinguish two generic categories classifying these functional forms: jet-based (\pt) or event-based (\ptj1, \htp). 
In the first case, in a given event, the scale used for the individual jet contributions is different for each jet while for an event-based scale prediction, a common scale is used for all jets in the event.

For a given fixed value of the central scale, the theoretical predictions for single jet inclusive observables, 
depend on the kinematics of the reconstructed jets, in particular on the 
jet cuts and the radius of the jet cone size $R$ used 
in the jet algorithm. 
The main difference between predictions obtained with 
different scale choices arise from events which are not in the Born $2\to 2$  back-to-back kinematical configuration. 
In this back-to-back kinematical situation, which is also reached at high $p_T$,
those scales ($\pt, \ptj1, \htp$), or multiples of these, are equivalent and using any of them yields the same predictions.

Away from these back-to-back configurations, the scales differ and 
besides the dominant leading jet contribution, the effect of the sub-leading jet contributions become sizeable. 
As a consequence, the impact of changing the scales becomes more and more important as the size of the jet cone $R$ decreases, 
where the importance of the sub-leading jet contributions is enhanced. 
To demonstrate this effect, we focus on two cone sizes  $R=0.4$ and $R=0.7$,
 as used by the LHC experimental collaborations.

Up to now, the most commonly used scale choices in describing single jet inclusive observables were \pt and \ptj1. 
In \cite{Cracowproc}, we showed that the NNLO predictions for these two scale choices differ substantially.
 While at high $p_T$,  a clear trend showing a stabilisation and agreement of the predictions is manifest, at low \pt, significant differences between the predictions persist and  are even more pronounced at NNLO. This 
 unexpected behaviour motivated the further study presented in \cite{jetscales}, where the different ingredients to 
 single jet inclusive distributions were analysed according to their behaviour at higher orders in the 
 perturbative expansion. 

\begin{figure*}[t]
  \centering
    \includegraphics[width=0.45\textwidth]{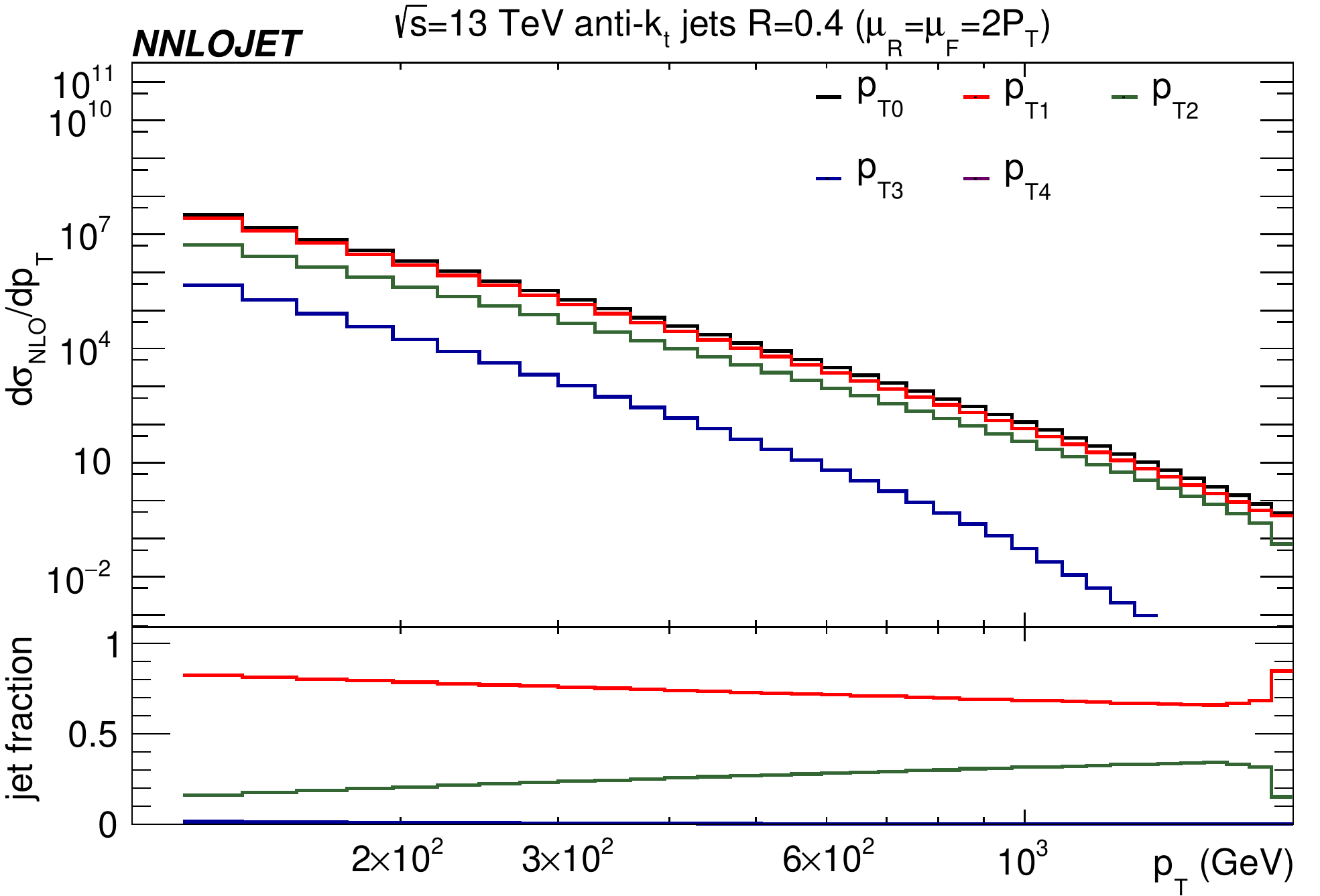}\hspace{1cm}
    \includegraphics[width=0.45\textwidth]{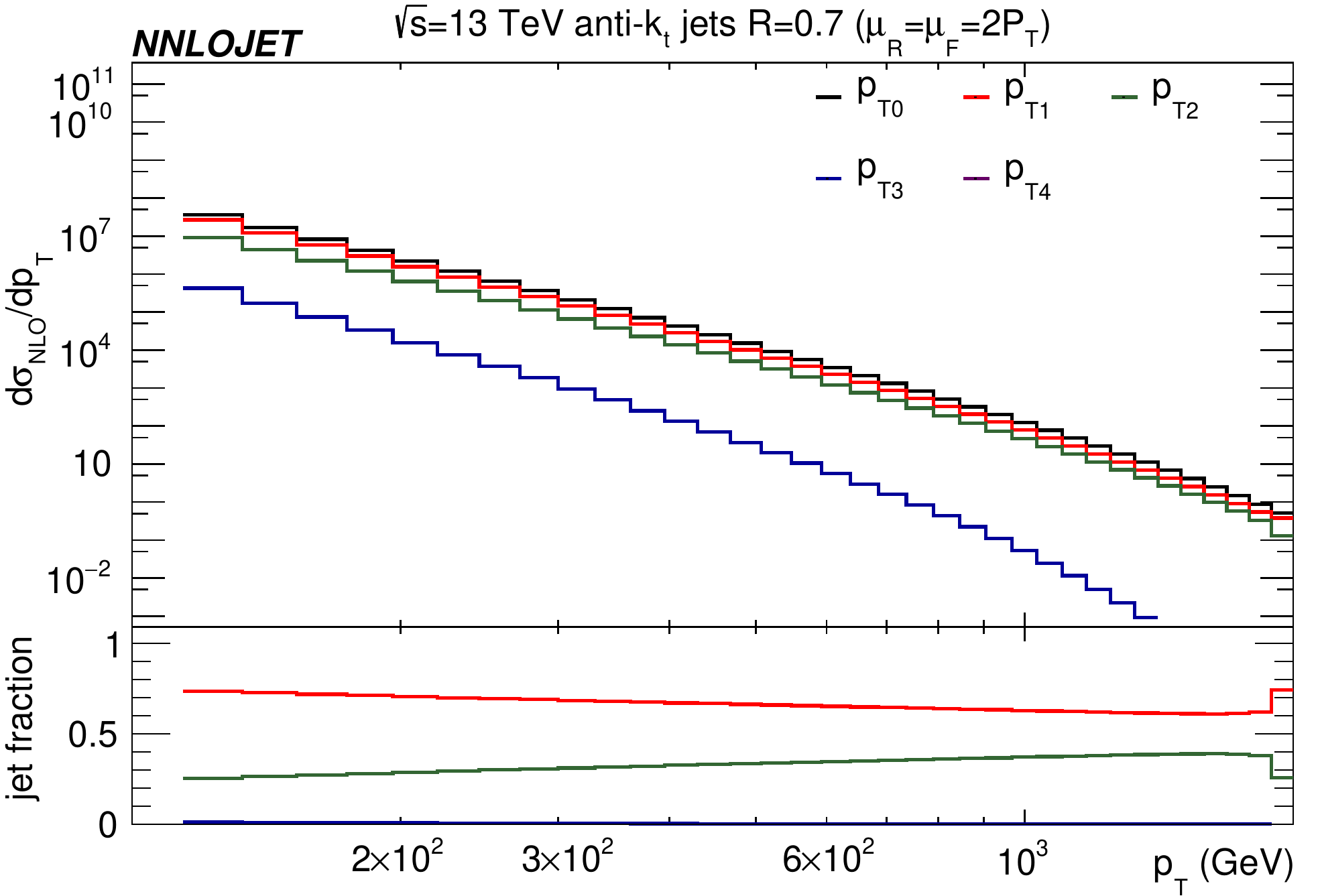}\\
    \includegraphics[width=0.45\textwidth]{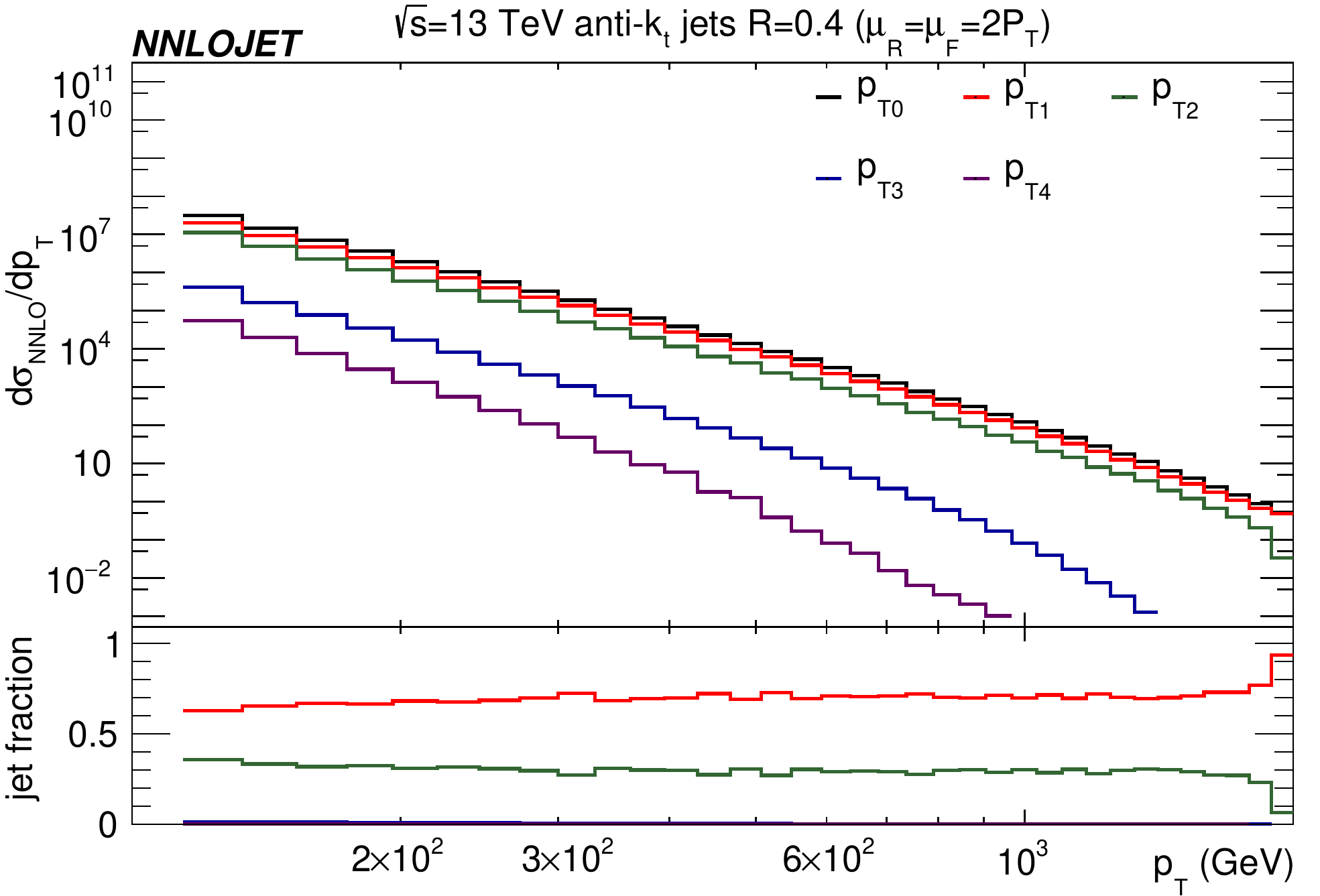}\hspace{1cm}
    \includegraphics[width=0.45\textwidth]{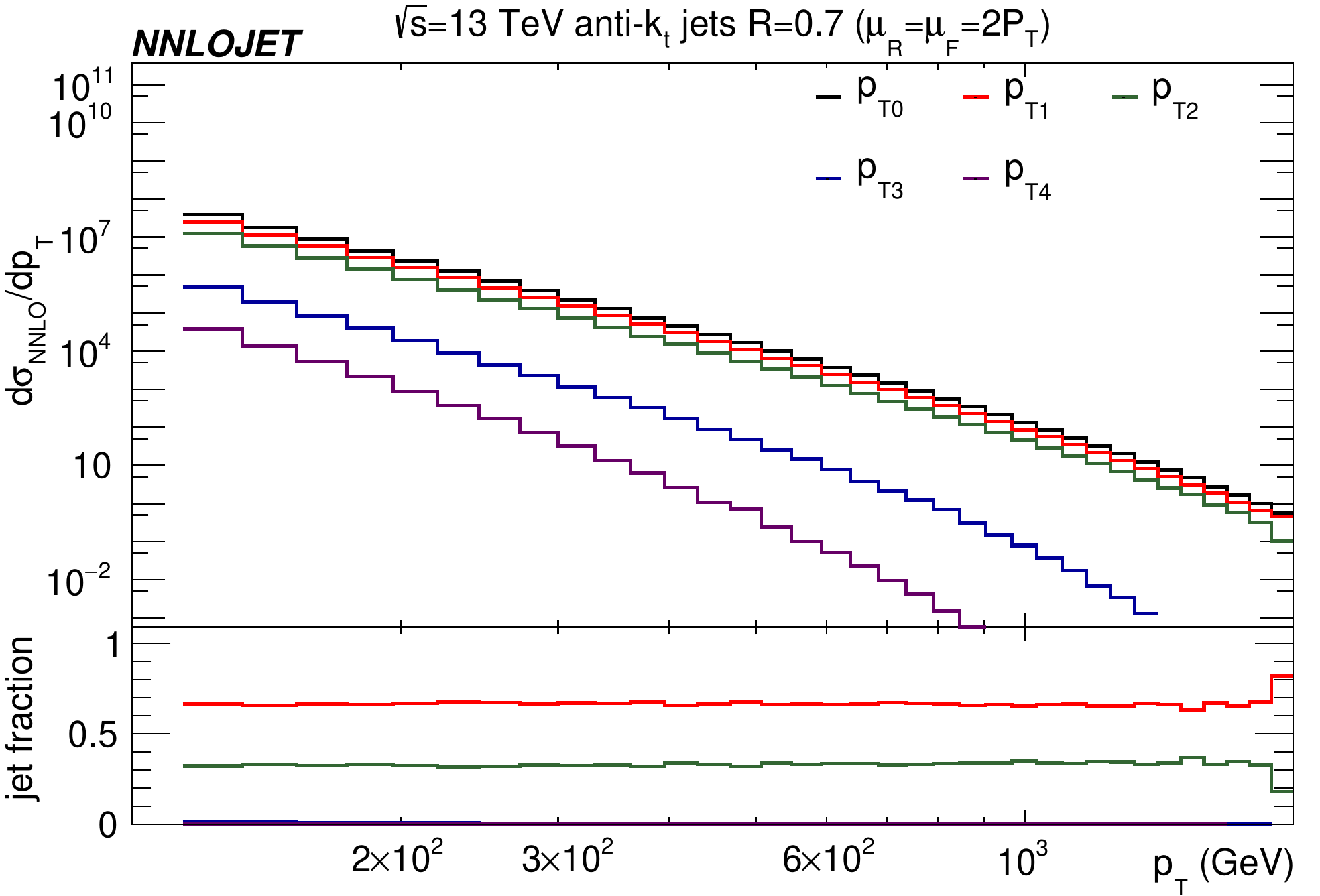}
\caption{Breakdown of the single jet inclusive cross section integrated over rapidity into contributions from first, second, third 
and fourth jet at NLO (upper plots) NNLO (lower plots) evaluated for $\mu=2 p_T$ for the jet cone sizes (left) $R=0.4$ and (right) $R=0.7$.
\label{fig:breakdown}}
\end{figure*}

It is worth mentioning that when decomposing the inclusive jet cross section in leading and sub-leading jet contributions, the individual 
jet distributions are well-defined and infrared safe only if they are inclusive in the jet rapidity. 
This is related to the fact that the rapidity assignment is not well-defined for the leading order kinematics where the transverse momentum of first and second jets are identical. When higher orders corrections are included, this can result in
 situations where the role of leading and sub-leading jet is interchanged between  event and counter-event, 
 thereby hampering their cancellation in infrared limits. 
In the inclusive jet transverse momentum distribution, this problem does not happen as all jets are summed over.

In order to be able to select the most appropriate scales for the theoretical description of the single jet inclusive production process, 
we define a list of desired properties, that a central scale choice 
 should satisfy in order to produce reliable predictions for single jet inclusive observables. 
 A detailed analysis 
of the leading and sub-leading jet distributions (in the context of a thorough comparison of \pt and \ptj1, see~\cite{jetscales}) 
 showed that some central scale choices  lead to infrared sensitive predictions with pathological behaviours 
regarding perturbative convergence and stability. It was found that the second jet distribution is particularly sensitive 
to the scale choice, sometimes even exhibiting an unphysical behaviour in predicting negative cross sections.

We therefore identify a set of requirements that a central scale choice should fulfil, 
 prior to any comparison with experimental data. 
These are: \emph{(a) perturbative convergence},
  \emph{(b) scale uncertainty as error estimate}, \emph{(c) perturbative convergence of the individual jet spectra}, 
  \emph{(d) stability of the second jet distribution}.

While the first two criteria are rather standard requirements, 
and have been used to select $m_{jj}$ as best choice for the central reference scale in di-jet production, 
the latter two criteria are specific to the single jet inclusive production and in particular to the single jet inclusive transverse momentum distribution. 

Using our selection procedure on the transverse momentum distributions integrated over rapidity and employing the CMS kinematical set-up \cite{CMS13TEV} to define the final state jets, we were able to identify $\mu=2\, \pt$ and $\mu=\hat{H}_{T}$ as the two theoretically best-motivated scale choices for single jet inclusive production. Note that the former belongs to the class of jet-based scales, the latter is an event-based scale and that the two scales coincide in Born kinematics.

\begin{figure*}[t]
  \centering
    \includegraphics[width=0.45\textwidth]{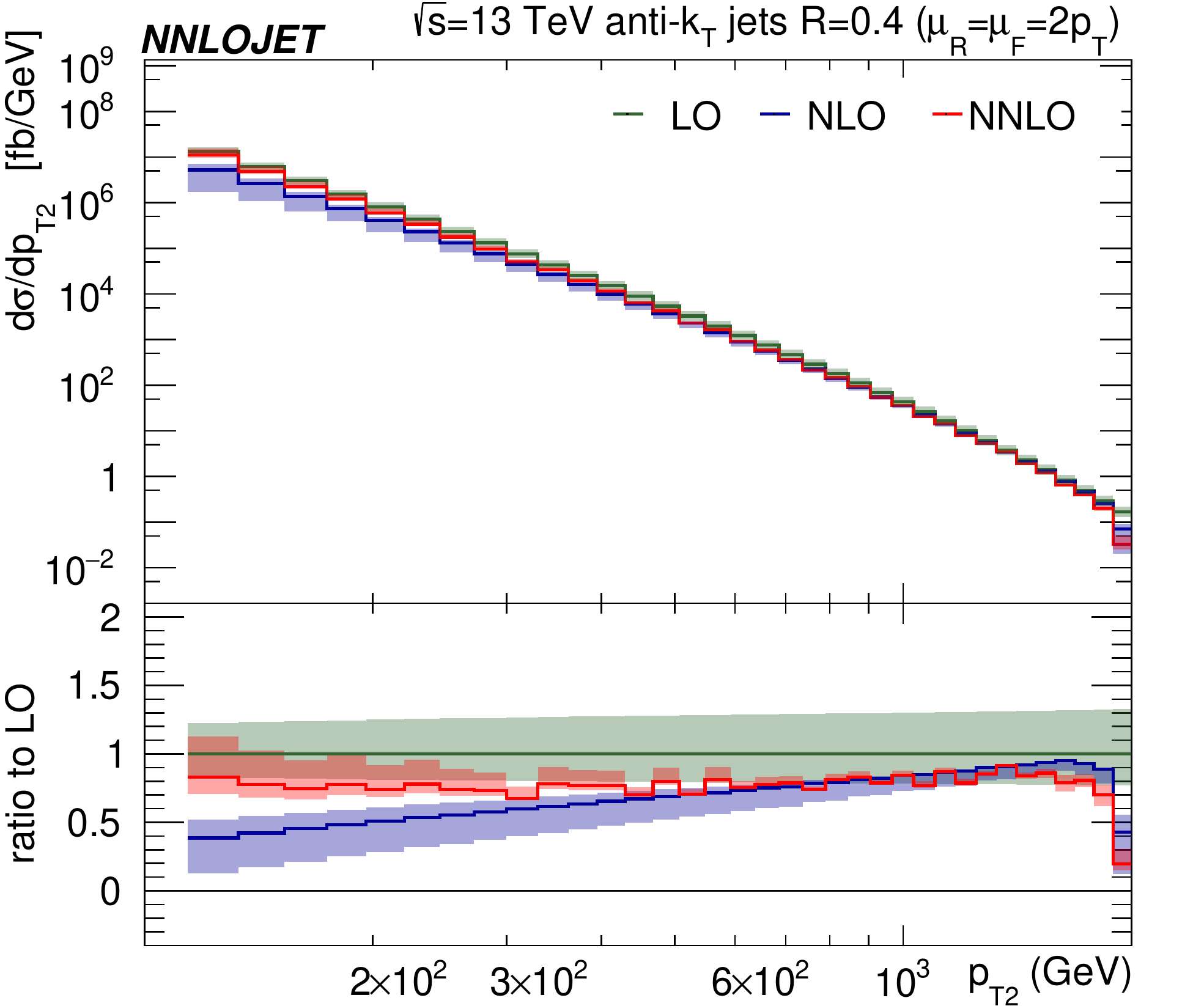}\hspace{1cm}
    \includegraphics[width=0.45\textwidth]{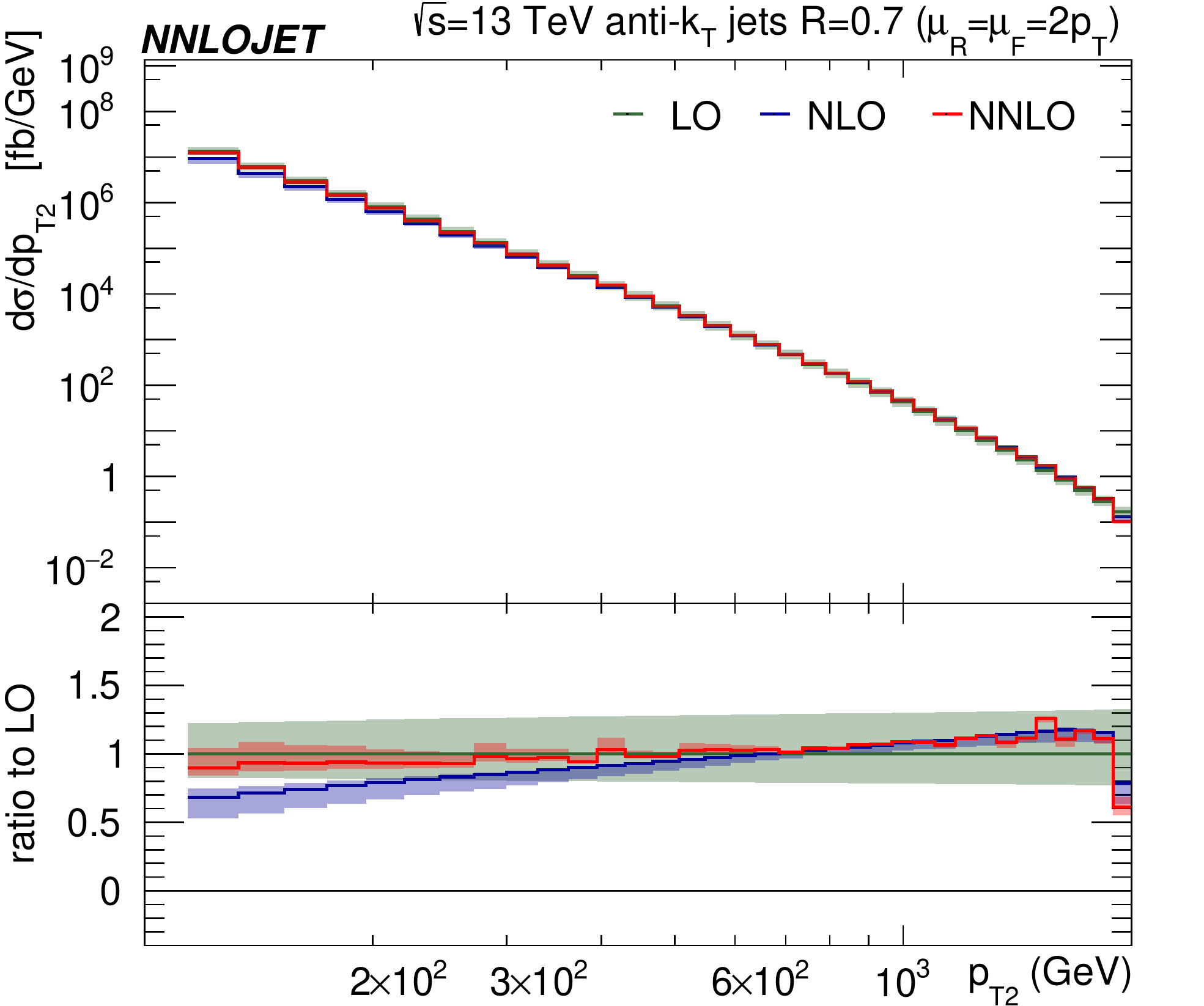}
\caption{Perturbative corrections to the transverse momentum  distribution of the second jet 
at 13 TeV (CMS cuts, $|y| < 4.7$, $R=0.4$ (left) and $R=0.7$ (right)),  integrated over rapidity and normalised to 
the LO prediction for the central scale choice $\mu=2\, \pt$. 
Shaded bands represent the theory uncertainty due to the variation of the factorization and renormalization scales.\label{fig:2ndptdist}}
\end{figure*}

In what follows, we present the most compelling features leading to 
this selection outcome for one of the selected central scale choices: $\mu=2\pt$. 
We start by presenting the breakdown of the single jet inclusive transverse momentum distribution into leading and sub-leading jet fractions. 
In Fig. \ref{fig:breakdown}, we see that over most of the $p_T$ range at NNLO, the leading jet contribution dominates, while 
the second jet fraction is sizeable and the third and fourth jet fractions are completely negligible for both cone sizes. 
At NLO, the second jet contribution becomes less and less important as $p_T$  decreases.
Going to NNLO, we observe a substantial increase 
in the second jet fraction compared to the NLO case specially at low $p_T$, 
the largest difference being observed for the smaller cone size $R=0.4$.

Given the potentially larger impact on the inclusive jet cross section of the second jet \pt distribution at NNLO, as compared 
to NLO,  we analysed its perturbative stability in further detail ~\cite{jetscales}.
In Fig.~\ref{fig:2ndptdist}, we observe that for $\mu=2\pt$, this distribution exhibits stable 
higher order corrections and small residual NNLO uncertainties while being positive over the whole $p_T$ 
range for both cone sizes. For most of the other central scale choices considered, these specific distributions 
yielded unphysical negative predictions.

\begin{figure*}[t]
\centering
  \includegraphics[width=0.45\textwidth]{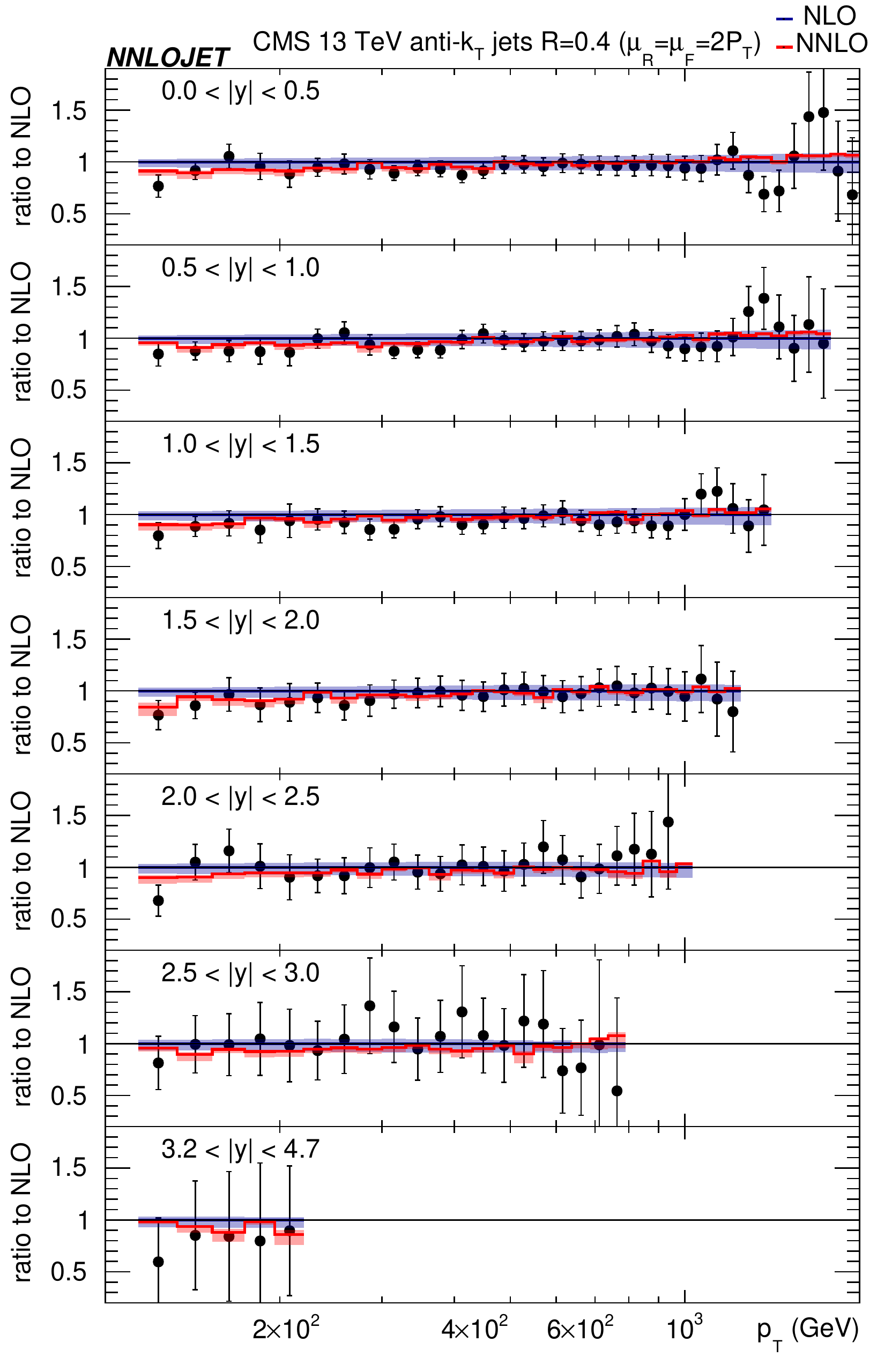} \hspace{1cm}
  \includegraphics[width=0.45\textwidth]{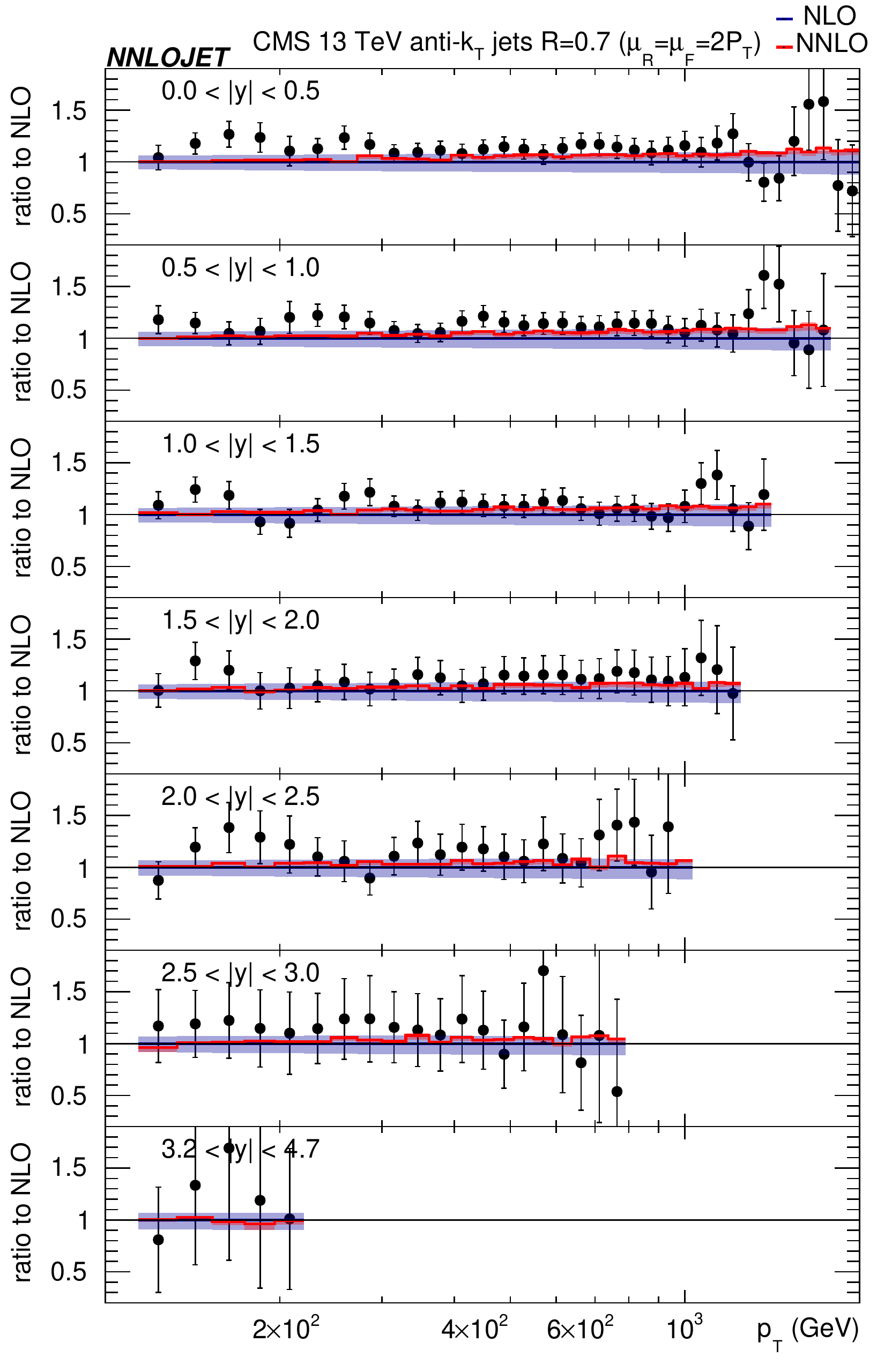}
  \caption{Double-differential single jet inclusive cross sections measurement by CMS~\cite{CMS13TEV} and NNLO perturbative QCD predictions as a function of the jet $\pt$ in slices of rapidity, for anti-$k_{T}$ jets with $R=0.4$ (left) and $R=0.7$ (right) normalised to the NLO 
   result $\mu=2\,\pt$, The shaded bands represent the scale uncertainty.
\label{fig:pTcompCMS}}
\end{figure*}

In \cite{jetscales}, the optimal scale choices $\mu=2\pt$ and $\mu=\htp$, have been validated further
by studying the distributions of the single jet cross section differential in transverse momentum and also in rapidity. 
Furthermore, for these scale choices, a direct comparison
for the double-differential single jet inclusive cross-sections to the CMS available data \cite{CMS13TEV} 
at $\sqrt{s}=13$~TeV has been performed.
As seen in Fig.~\ref{fig:pTcompCMS} for $\mu=2\pt$, we observe small positive NNLO corrections across all rapidity slices,
that improve the agreement with the CMS data, as compared to the NLO prediction for both cone sizes. 
In addition, across the entire $\pt$ range, a clear reduction of the scale uncertainty is manifest when going from NLO to NNLO.  
 
\section{ Conclusions and outlook}
In this talk, we have highlighted the main outcomes of recent studies regarding the choice of a reference central scale for the factorisation and renormalisation scales in observables related to the di-jet and single jet inclusive production processes~\cite{dijets,jetscales}. 
We have emphasized the substantial differences between the perturbative behaviour of their differential cross sections.
While in the di-jet case, the knowledge of the NNLO corrections was sufficient to establish $m_{jj}$ as the preferred choice based on standard perturbative convergence and stability criteria, we saw that in the case of the $\pt$ spectrum in single jet inclusive production a more elaborated list of requirements is needed to identify  
$\mu=2\,\pt$ and $\mu=\htp$ as the most appropriate 
from a list of a priori equally valid and reasonable scale choices.
We saw in particular that $\mu=2\pt$ fulfils the selection criteria associated to the second jet contribution, whose impact 
in the single jet inclusive cross section is particularly sensitive to the instabilities. 
Using these theoretically well-motivated scale choices, the NNLO predictions for di-jet invariant 
mass and single jet $\pt$ distribution 
are in good agreement with the data  presenting a significant reduction of the scale 
uncertainty over most of the allowed kinematical range in $m_{jj}$ and $\pt$ respectively. 
We expect that these results will enable precision phenomenology with  jet data, such as the NNLO
determination of the parton distributions functions and of fundamental QCD parameters. 

\acknowledgments 
The authors thank Xuan Chen, Juan Cruz-Martinez, Rhorry Gauld, Marius H\"ofer, Imre Majer, Tom Morgan, Jan Niehues, Duncan Walker and James Whitehead for useful discussions and their many contributions to the \textsc{NNLOjet} code.
This research was supported in part by the UK Science and Technology Facilities Council, by the Swiss National Science Foundation (SNF) under contracts 200020-175595 and 200021-172478, and CRSII2-160814, by the Research Executive Agency (REA) of the European Union through the ERC Advanced Grant MC@NNLO (340983) and by the Funda\c{c}\~{a}o para a Ci\^{e}ncia e Tecnologia (FCT-Portugal), project
UID/FIS/00777/2013.

\end{document}